\documentclass[10pt]{article}
\usepackage{amsmath, amsthm}

\newtheoremstyle{theorem}
{10pt} 
{10pt} 
{\sl} 
{\parindent} 
{\bf} 
{. } 
{ } 
{} 
\theoremstyle{theorem}

\newtheoremstyle{defi}
{10pt} 
{10pt} 
{\rm} 
{\parindent} 
{\bf} 
{. } 
{ } 
{} 
\theoremstyle{defi}

\begin{document}

\title{Influence of probability
density function of the passage time in the method of
non-equilibrium statistical operator on non-equilibrium properties
of the system}

\author{V.V. Ryazanov\\
Institute for Nuclear Research, Kiev, pr.Nauki, 47 Ukraine\\
vryazan@kinr.kiev.ua}

\maketitle

\begin{abstract}
A family of non-equilibrium statistical operators (NSO) is
introduced which differ by the system lifetime distribution over
which the quasi-equilibrium (relevant) distribution is averaged.
This changes the form of the source in the Liouville equation, as
well as the expressions for the kinetic coefficients, average
fluxes, and kinetic equations obtained with use of NSO. It is
possible to choose a class of lifetime distributions for which
thermodynamic limiting transition and to tend to infinity of
average lifetime of system is reduced to the result received at
exponential distribution for lifetime, used by Zubarev. However
there is also other extensive class of realistic distributions of
lifetime of system for which and after to approach to infinity of
average lifetime of system non-equilibrium properties essentially
change. For some distributions the effect of "finite memory" when
only the limited interval of the past influence on behaviour of
system is observed. It is shown, how it is possible to spend
specification the description of effects of memory within the
limits of NSO method, more detailed account of influence on
evolution of system of quickly varying variables through the
specified and expanded form of density of function of distribution
of lifetime. The account of character of history of the system,
features of its behaviour in the past, can have substantial
influence on non-equilibrium conduct of the system in a present
moment time.\\

{\bf AMS Subject Classification: 82C03; 82C70}\\

{\bf Key Words and Phrases: non-equilibrium statistical operator,
lifetime, account of character of history of the system}
\end{abstract}

\section{Introduction}

One of the most fruitful and successful ways of development of the
description of the non-equilibrium phenomena are served by a
method of the non-equilibrium statistical operator (NSO)
\cite{zub71,zub80,kal,kal71,zub96}. In work \cite{ry01} new
interpretation of a method of the NSO is given, in which operation
of taking of invariant part \cite{zub71,zub80,kal,kal71,zub96} or
use auxiliary "weight function" (in terminology \cite{ra95,ra99})
in NSO are treated as averaging of quasi-equilibrium statistical
operator on distribution of past lifetime of system. In work
\cite{zub80} is chosen uniform distribution for an initial moment
$t_{0}$, which after change of integration passes to the
exponential distribution $p_{q}(u)= exp\{-\varepsilon u\}$. Such
distribution serves as the limiting distribution of lifetime
\cite{str61}, of the first passage time of level. In general case
it is possible to choose a lot of functions for the obvious type
of distribution $p_{q}(u)$, that marked in works \cite{ry01,ry07}.
This approach adjust with the operations spent in the general
theory of random processes, in the renewal theory, and also with
the lead Zubarev in work as \cite{zub80} reception NSO by means of
averaging on the initial moment of time.

In Kirkwood's works \cite{kirk} it was noticed, that the system
state in time present situation depends on all previous evolution
of the non-equilibrium processes developing it. For example, in
real crystals it is held in remembrance their formation in various
sorts "defects", reflected in structure of the crystals. Changing
conditions of formation of crystals, we can change their
properties and create new materials. In works \cite{ra95,ra99} it
is specified, that it is possible to use many "weight functions".
Any form of density of lifetime distribution gives a chance to
write down a source of general view in dynamic Liouville equation
which thus becomes, specified Boltzmann and Prigogine
\cite{ra95,ra99,Prig}, and contains dissipative items.

If in Zubarev's works \cite{zub71,zub80} the linear form of a
source corresponding limiting exponential distribution for
lifetime is used other expressions for density of lifetime
distribution give fuller and exact analogues of "integrals of
collisions".

In works \cite{rep83,der85} sources in the Liouville equation
different from sources, entered in the NSO method in works
\cite{zub71,zub80} are considered. But in works \cite{rep83,der85}
at $\varepsilon\rightarrow 0$ (and $\Gamma\rightarrow \infty$ in
interpretation paper \cite{ry01}) this source turn into a zero. In
the present work the sources which are not turn into a zero at
$\Gamma\rightarrow \infty$ are considered. Such source is caused
by external influences on system. The approach of present paper
differs from the methods used in \cite{rep83,der85}. But use of
distribution of a lifetime of system in the present work can be
compared with noted in work \cite{rep83} enlarged the set of
macroobservables, include besides the physically necessary
macroobservables additional ones, namely lifetime.

In work \cite{rau} irreversible transfer equations are received in
assumptions of coarsening of the distributions, a certain choice
of macroscopical variables and the analysis of division of time
scales of the description (last circumstance was marked in
\cite{bog}). Importance and necessity of the analysis of the time
scales playing a fundamental role in the description of
macroscopical dynamics of system is underlined. Evolution of slow
degrees of freedom is described by Markovian equations. Thus the
time scale on which observable variables evolve, should be much
more time of memory on which the residual effects brought by
irrelevant degrees of freedom are considered. As marked in paper
\cite{der85}, it requires  the condition of a maximum damping of
the non-macroscopic information.

Otherwise effects of memory play an essential role. Memory time is
estimated in work \cite{rau} for Boltzmann equation. In the
present work the consideration subject is made by situations when
it is necessary to consider effects of memory. Examples of such
situations are given in \cite{rau}.

In work \cite{ry07} it is shown, in what consequences for
non-equilibrium properties of system results change of lifetime
distribution of system for systems of the limited volume with
finite lifetime. In the present work are considered also
infinitely greater systems with infinite average lifetime.

\section{New interpretation of NSO}
\label{sect:1}

In \cite{ry01} the Nonequilibrium Statistical Operator introduced
by Zubarev \cite{zub71,zub80} rewritten as

\begin{equation}
ln\varrho(t)= \int_{0}^{\infty}p_{q}(u)ln\varrho_{q}(t-u,
-u)du,\quad
ln\varrho_{q}(t, 0)=-\Phi(t)-\sum_{n}F_{n}(t)P_{n};\\
$$
$$
ln\varrho_{q}(t, t_{1})=e^{\{-t_{1}H/i\hbar\}}ln\varrho_{q}(t,
0)e^{\{t_{1}H/i\hbar\}}; \quad \Phi(t)=\ln Sp
\exp\{\sum_{n}F_{n}(t)P_{n}\}, \label{zu}
\end{equation}
where $H$ is hamiltonian, $ln\varrho(t)$ is the logarithm of the
NSO in Zubarev's form, $ln\varrho_{q}(t, 0)$ is the logarithm of
the quasi-equilibrium (or relevant); the first time argument
indicates the time dependence of the values of the thermodynamic
parameters $F_{m}$; the second time argument $t_{2}$ in
$\varrho_{q}(t_{1}, t_{2})$ denotes the time dependence through
the Heizenberg representation for dynamical variables $P_{m}$ from
which $\varrho_{q}(t, 0)$ can depend
\cite{zub71,zub80,ry01,ra95,ra99}. In \cite{ry01} the auxiliary
weight function $p_{q}(u)=\varepsilon exp\{-\varepsilon u\}$ was
interpreted as the density of probability distribution of lifetime
$\Gamma$ of a system. There $\Gamma$ is random variables of
lifetime from the moment $t_{0}$ of its birth till the current
moment $t$; $\varepsilon^{-1}=\langle t-t_{0}\rangle$; $\langle
t-t_{0}\rangle=\langle\Gamma\rangle$, where
$\langle\Gamma\rangle=\int u p_{q}(u)du$ is average lifetime of
the system. This time period can be called the time period of
getting information about system from its past. Instead of the
exponential distribution $p_{q}(u)$ in (1) any other sample
distribution could be taken. This fact was marked in \cite{ry01}
and \cite{ra95,ra99} (where the distribution density $p_{q}(u)$ is
called auxiliary weight function $w(t,t')$). From the complete
group of solutions of Liouville equation (symmetric in time) the
subset of retarded "unilateral" in time solutions is selected by
means of introducing a source in the Liouville equation for
$\ln\varrho(t)$

\begin{equation}
\frac{\partial\ln\varrho(t)}{\partial
t}+iL\ln\varrho(t)=-\varepsilon(\ln\varrho(t)-\ln\varrho_{q}(t,
0))=J, \label{li}
\end{equation}
which tends to zero (value $\varepsilon\rightarrow 0$) after
thermodynamic limiting transition. Here $L$ is Liouville operator;
$iL=-\{H,\varrho\}=\Sigma_{k}[\frac{\partial H}{\partial
p_{k}}\frac{\partial\varrho}{\partial q_{k}}-\frac{\partial
H}{\partial q_{k}}\frac{\partial\varrho}{\partial p_{k}}]$; $H$ is
Hamilton function, $p_{k}$ and $q_{k}$ are momenta and coordinates
of particles; $\{...\}$ is Poisson bracket. In \cite{mor} it was
noted that the role of the form of the source term in the
Liouville equation in NSO method has never been investigated. In
\cite{fe} it is stated that the exponential distribution is the
only one which possesses the Markovian property of the absence of
contagion, that is whatever is the actual age of a system, the
remaining time does not depend on the past and has the same
distribution as the lifetime itself. It is known
\cite{zub71,zub80,ry01,ra95,ra99} that the Liouville equation for
NSO contains the source $J=J_{zub}=-\varepsilon [ln\varrho
(t)-ln\varrho_{q}(t,0)]$ which becomes vanishingly small after
taking the thermodynamic limit and setting $\varepsilon\rightarrow
0$, which in the spirit of the paper \cite{zub71} corresponds to
the infinitely large lifetime value of an infinitely large system.
For a system with finite size this source is not equal to zero. In
\cite{ra99} this term enters the modified Liouville operator and
coincides with the form of Liouville equation suggested by
Prigogine \cite{Prig} (the Boltzmann-Prigogine symmetry), when the
irreversibility is entered in the theory on the microscopic level.
We note that the form of NSO by Zubarev cast in \cite{ry01}
corresponds to the main idea of \cite{Prig} in which one sets to
the distribution function $\varrho$ ($\varrho_{q}$ in Zubarev's
approach) which evolves according to the classical mechanics laws,
the coarse distribution function $\widetilde{\varrho}$
($\varrho(t)$ in the case of Zubarev's NSO) whose evolution is
described probabilistically since one perform an averaging with
the probability density $p_{q}(u)$. The same approach (but instead
of the time averaging the spatial averaging was taken) was
performed in \cite{klim}.

Besides the Zubarev's form of NSO \cite{zub71,zub80}, NSO
Green-Mori form \cite{gr,mo} is known, where one assumes the
auxiliary weight function \cite{ra95} to be equal
$W(t,t')=1-(t-t')/\tau; w(t,t')=dW(t,t')/dt'=1/\tau;
\tau=t-t_{0}$. After averaging one sets $\tau\rightarrow\infty$.
This situation at $p_{q}(u=t-t_{0})=w(t,t'=t_{0})$ coincides with
the uniform lifetime distribution. The source in the Liouville
equation takes the form $J=ln\varrho_{q}/\tau$. In \cite{zub71}
this form of NSO is compared to the Zubarev's form.

One could name many examples of explicit defining of the function
$p_{q}(u)$. Every definition implies some specific form of the
source term $J$ in the Liouville equation, some specific form of
the modified Liouville operator and NSO. Thus the family of NSO is
defined.

\section{Modifications to the nonequilibrium description}
\label{sect:2}

Let's consider now, what consequences follow from such
interpretation of NSO.

\subsection{Families of NSO}
\label{subsect:A}

Setting various distributions for past lifetime of the system, we
receive a way of recording of families of NSO. Class of NSO from
this family will be connected with a class of distributions for
lifetime (taken, for example, from the stochastic theory of
storage processes, the theory of queues etc.) and with relaxation
properties of that class of physical systems which is
investigated. The general expression for NSO with any distribution

\begin{equation}
ln\varrho(t)= \int_{0}^{\infty}p_{q}(u)ln\varrho_{q}(t-u,
-u)du=
\label{NSO}
\end{equation}

$$
=ln\varrho_{q}(t, 0)-\int_{0}^{\infty}(\int p_{q}(u)du)\frac{d
\ln\varrho(t-u, -u)}{du}du,
$$
where integration by parts in time is carried out at  $\int p_{q}
(y)dy_{|y=0}=-1; \\ \int p_{q}(y)dy_{|y\rightarrow\infty}=0$; at
$p_{q}(y)=\varepsilon\exp\{-\varepsilon y\};
\varepsilon=1/\langle\Gamma\rangle$, the expression (1) passes in
NSO from \cite{zub71,zub80}. In \cite{fe} it is shown, how from
random process ${X (t)}$, corresponding to evolution of
quasi-equilibrium system, it is possible to construct set of new
processes, introducing the randomized operational time. It is
supposed, that to each value $t>0$ there corresponds a random
value $\Gamma(t)$ with the distribution $p^{t}_{q}(y)$. The new
stochastic kernel of distribution of a random variable
$X(\Gamma(t))$ is defined by equality of a kind (1). Random
variables $X(\Gamma(t))$ form new random process which, generally
speaking, need not to be of Markovian type any more. Each moment
of time $t$ of "frozen" quasi-equilibrium system is considered as
a random variable $\Gamma(t)$ the termination of lifetime with
distribution $p^{t}_{q}(y)$. Any moment of lifetime can be with
certain probability the last. That the interval $t-t_{0}=y$ was
enough large (that became insignificant details of an initial
condition as dependence on the initial moment $t_{0}$ is
nonphysical \cite{zub71,zub80}), it is possible to introduce the
minimal lifetime $\Gamma_{min}=\Gamma_{1}$ and to integrate in
(\ref{NSO}) on an interval $(\Gamma_{1},\infty)$. It results to
the change of the normalization density of distribution
$p_{q}(y)$. For example, the function $p_{q}(y)=\varepsilon\exp
\{-\varepsilon y\}$ will be replaced by $p_{q}(y)=\varepsilon\exp
\{\varepsilon\Gamma_{1}-\varepsilon y\}, y\geq\Gamma_{1};
p_{q}(y)=0, y < \Gamma_{1}$. The under limit of integration in
(\ref{NSO}) by $\Gamma_{1}\rightarrow 0$ is equal $0$. It is
possible to choose $p_{q}(y)=Cf(y), y<t_{1};  p_{q}(y)
=\varepsilon\exp \{-\varepsilon y\}, y\geq t_{1};
C=(1-exp\{-\varepsilon t_{1}\})/(\int_{0}^{t_{1}} f(y)dy)$. The
function $f(y)$ can be taken from models of the theory of queues,
the stochastic theory of storage and other sources estimating the
lifetime distribution for small times (for example
\cite{co,cox,str61,tur}). The value $t_{1}$ can be found from
results of work \cite{str61}. It is possible to specify many
concrete expressions for lifetime distribution of system, each of
which possesses own advantages. To each of these expressions there
corresponds own form of a source in Liouville equation for the
nonequilibrium statistical operator. In general case any functions
$p_{q}(u)$ the source is:

\begin{equation}
J=p_{q}(0)\ln\varrho_{q}(t, 0)+\int_{0}^{\infty}\frac{\partial
p_{q}(y)}{\partial y}(ln\varrho_{q}(t-y, -y))dy \label{sour}
\end{equation}
(when values $p_{q}(0)$ disperse, it is necessary to choose the
under limit of integration equal not to zero, and $\Gamma_{min}$).
Such approach corresponds to the form of dynamic Liouville
equation in the form of Boltzmann-Bogoliubov-Prigogine
\cite{ra95,ra99,Prig}, containing dissipative items.

Thus the operations of taking of invariant part \cite{zub71},
averaging on initial conditions \cite{zub80}, temporary
coarse-graining \cite{kirk}, choose of the direction of time
\cite{ra95,ra99}, are replaced by averaging on lifetime
distribution.

It is essentially that and in exponential distribution from
\cite{zub71,zub80} $\varepsilon\neq 0$. The thermodynamic limiting
transition is not performed, and actually important for many
physical phenomena dependence on the size of system are
considered. We assume $\varepsilon$ and $\langle\Gamma\rangle$ to
be finite values. Thus the Liouville equation for $\varrho(t)$
contains a finite source. The assumption about finiteness of
lifetime breaks temporary symmetry. And such approach
(introduction $p_{q}(y)$, averaging on it) can be considered as
completing the description of works \cite{zub71,zub80}.

In work \cite{str61} lifetimes of system are considered as the
achievement moments by the random process characterizing system,
certain border, for example, zero. In \cite{str61} are received
approached exponential expressions for density of probability of
lifetime, accuracy of these expressions is estimated. In works
\cite{str95,str96} lifetimes of molecules are investigated, the
affinity of real distribution for lifetime and approached
exponential model is shown. It is possible to specify and other
works (for example \cite{gasp93,gasp95,dorf}) where physical
appendices of concept of lifetime widely applied in such
mathematical disciplines, as reliability theory, the theory of
queues and so forth (under names non-failure operation time, the
employment period, etc.) are considered. In the present section
lifetime joins in a circle of the general physical values, acting
in an estimation or management role (in terminology of the theory
of the information \cite{str66}) for the quasi-equilibrium
statistical operator that allows to receive the additional
information on system.

Let's notice, that in a case when value
$d\ln\varrho_{q}(t-y,-y)/dy$ (the operator of entropy production
$\sigma$ \cite{zub71}) in the second item of the right part
(\ref{NSO}) does not depend from $y$ and is taken out from under
integral on $y$, this second item becomes
$\sigma\langle\Gamma\rangle$, and expression (\ref{NSO}) does not
depend on form of function $p_{q}(y)$. There is it, for example,
at $\varrho_{q}(t)\sim \exp\{-\sigma t\}, \sigma=const$. In work
\cite{dew} such distribution is received from a principle of a
maximum of entropy at the set of average values of fluxes.

\subsection{Physical sense of distributions for past
lifetime of system} \label{subsect:B}

As is known (for example, \cite{co}, \cite{str61}, \cite{tur}),
exponential distribution for lifetime

\begin{equation}
p_{q}(y) =\varepsilon\exp\{-\varepsilon  y\}, \label{expDi}
\end{equation}
used in Zubarev's works \cite{zub71,zub80}, is limiting
distribution for lifetime, fair for large times. It is marked in
works \cite{zub71,zub80} where necessity of use of large times
connected with damping of nonphysical initial correlations. Thus,
in works \cite{zub71,zub80} the thermodynamic result limiting and
universal is received, fair for all systems. It is true in a
thermodynamic limit, for infinitely large systems. However real
systems have the finite sizes. Therefore essential there is use of
other, more exact distributions for lifetime. In this case the
unambiguity of the description peculiar to a thermodynamic limit
is lost \cite{mart}.

For NSO with Zubarev's function (\ref{expDi}) the value enter in
second item

\begin{equation}
-\int p_{q}(y)dy =\exp\{-\varepsilon  y\}=1-\varepsilon
y+(\varepsilon
y)^{2}/2-...=1-y/\langle\Gamma\rangle+y^{2}/2\langle\Gamma\rangle^{2}-...
\label{sum}
\end{equation}
Obviously that to tend to infinity of average lifetime,
$\langle\Gamma\rangle\rightarrow\infty$, correlation (\ref{sum})
tends to unity.

Besides exponential density of probability (\ref{expDi}), as
density of lifetime distribution Erlang distributions (special or
the general), gamma distributions etc. (see \cite{co,cox}), and
also the modifications considering subsequent composed asymptotic
of the decomposition \cite{tur} can be used. General Erlang
distributions for $n$ classes of ergotic states are fair for cases
of phase transitions or bifurcations. For $n=2$ general Erlang
distribution looks like
$p_{q}(y)=\theta\rho_{1}\exp\{-\rho_{1}y\}+(1-\theta)\rho_{2}\exp\{-\rho_{2}y\},
\theta<1$. Gamma distributions describe the systems which
evolution has some stages (number of these stages coincides with
gamma distribution order). Considering real-life stages in
non-equilibrium systems (chaotic, kinetic, hydrodynamic, diffusive
and so forth \cite{bog}), it is easy to agree, first, with
necessity of use of gamma distributions of a kind

\begin{equation}
p_{q}(y) =\varepsilon(\varepsilon y)^{k-1}\exp\{-\varepsilon
y\}/\Gamma(k) \label{gam}
\end{equation}
($\Gamma(k)$ is gamma function, at $k=1$ we receive distribution
(\ref{expDi})), and, secondly, - with their importance in the
description of non-equilibrium properties. The
piecewise-continuous distributions corresponding to the different
stages of evolution of the system will be used below.

More detailed description $p_{q}(u)$ in comparison with limiting
exponential (\ref{expDi}) allows to describe more in detail real
stages of evolution of system (and also systems with small
lifetimes). Each from lifetime distributions has certain physical
sense. In the theory of queues, for example \cite{prab}, to
various disciplines of service there correspond various
expressions for density of lifetime distribution. In the
stochastic theory of storage \cite{prab}, to these expressions
there correspond various models of an exit and an input in system.

The value $\varepsilon$ without taking into account of dissipative
effects can be defined, for example, from results of work
\cite{str61}. The value $\varepsilon$ is defined also in work
\cite{ry01} through average values of operators of entropy and
entropy production, flows of entropy and their combination.

How was already marked, it is possible to specify very much, no
less than $1000$, expressions for the distributions of past
lifetime of the system. Certain physical sense is given to each of
these distributions. To some class functions of distributions,
apparently, some class of the physical systems corresponds, the
laws of relaxation in which answer this class of functions of
distributions for lifetime.

\subsection{Influence of the past on
non-equilibrium properties} \label{subsect:C}

In \cite{fe} by consideration of the paradox connected with a
waiting time, the following result is received: let $X_{1}=S_{1};
X_{2}=S_{2}-S_{1}; ...$ are mutually independent also it is
equally exponential the distributed values with average
$1/\varepsilon$. Let $t>0$ is settled, but it is any. Element
$X_{k}$, satisfying to condition $S_{k-1}<t  \leq S_{k}$, has
density $\nu_{t}(x)=\varepsilon^{2}x\exp\{-\varepsilon x\}, 0<x
\leq t; \nu_{t}(x)=\varepsilon(1+\varepsilon x)\exp\{-\varepsilon
x\}, x>t$. In Zubarev's NSO \cite{zub71,zub80} the value of
lifetime to a present moment t, belonging lifetime $X_{k}$,
influence of the past on the present is considered. Therefore the
value $p_{q}(u)$ should be chosen not in the form of exponential
distribution (\ref{expDi}), and in a form

\begin{equation}
p_{q}(y) =\varepsilon^{2}y\exp\{-\varepsilon  y\}, \label{exp2}
\end{equation}
that in the form of gamma distribution (\ref{gam}) at $k=2$. In
this case distribution (\ref{exp2}) coincides with special Erlang
distribution of order $2$ \cite{co}, when refusal (in this case -
the moment $t$) comes in the end of the second stage \cite{co},
the system past consists of two independent stages. Function of
distribution is equal $P_{q}(x)=1-\exp\{-\varepsilon
x\}-\varepsilon x\exp\{-\varepsilon x\}, p_{q}(u)=dP_{q}(u)/du$,
unlike exponential distribution, when
$P_{q}(x)=1-\exp\{-\varepsilon x\}$. The behaviour of these two
densities of distribution of a form (\ref{expDi}) and (\ref{exp2})
essentially differs in a zero vicinity. In case of (\ref{exp2}) at
system low probability to be lost at small values $y$, unlike
exponential distribution (\ref{expDi}) where this probability is
maximal. Any system if has arisen, exists any minimal time, and it
is reflected in distribution (\ref{exp2}).

In work \cite{ry07} it is shown, in what differences from
Zubarev's distribution (\ref{NSO}) with exponential distribution
of lifetime (\ref{expDi}) results gamma distribution (\ref{gam}),
(\ref{exp2}) use. Additional items in NSO, in integral of
collisions of the generalized kinetic equation, in expressions for
average fluxes and self-diffusions coefficient are considered. The
same in [10] is done and for special Erlang distribution
$k=2,3,4..., n, P_{q}(x)=1-\exp\{-\varepsilon x\}[1+\varepsilon
x/1! + ... +(\varepsilon x)^{k-1}/(k-1)!];
\varepsilon=k/\langle\Gamma\rangle$. For distributions
(\ref{gam}), (\ref{exp2}) is correct correlation (4), value $-\int
p_{q}(u)du \rightarrow 1$ by
$\langle\Gamma\rangle\rightarrow\infty$.

Thus the multi-stage model of the past of system is introduced.
Non-equilibrium processes usually proceed in some stages, each of
which is characterized by the time scale. In distribution
(\ref{exp2}) the account of two stages, possibly, their minimal
possible number is made. Other distributions can describe any
other features of the past. Corresponding additives will be
included into expressions for fluxes, integral of collisions,
kinetic coefficients. Besides special Erlang distributions with
whole and specified values $k=n$, that does not deduce us from set
of one-parametrical distributions, the general already
two-parametrical gamma distribution where the parameter $k$ can
accept any values can be used. In this case $\langle\Gamma\rangle\
=k/\varepsilon$. The situation (formally), when $k<1$ is possible.
Then sources will tends to infinity, as
$(t-t_{0})^{k-1}_{|t\rightarrow t_{0}}\rightarrow\infty$ at $k<1$.
This divergence can be eliminated, having limited from below the
value $t-t_{0}$ of minimal lifetime value $\Gamma_{min}$, having
replaced the under zero limit of integration on $\Gamma_{min}$.
Then to expression for a source (\ref{sour}) it is added item
$[(\varepsilon
\Gamma_{min})^{k-1}/\Gamma(k)]\varepsilon\exp\{-\varepsilon
\Gamma_{min}\}\ln\varrho_{q}(t - \Gamma_{min}, - \Gamma_{min})$.

In \cite{fe} it was shown that the exponential lifetime
distribution $t_{f}-t_{0}$ ($t_{f}$ , $t_{0}$ are random moments
of system death and birth) at big $t$'s the "age" of a system
$t-t_{0}$ tends to the exponential form. In Zubarev's NSO
\cite{zub71,zub80} the lifetime value $t-t_{0}$ to the current
time $t$, which is a part of the total lifetime $t_{f}-t_{0}$, is
considered, that is the influence of the past on the current
moment is taken into account. The full lifetime distribution, as
well as the "past" lifetime (i.e. time from the system birth
$t_{0}$ till the current time $t$) need not be exponential. The
logarithm of NSO in the case (\ref{exp2}) has the form

\begin{equation}
ln\varrho(t)= \int_{0}^{\infty}p_{q}(u)\ln\varrho_{q}(t-u,-u)du=
\label{exp2n}
\end{equation}
$$
\int_{0}^{\infty}\varepsilon^{2}u\exp\{-\varepsilon
u\}\ln\varrho_{q}(t-u,-u)du=
$$
$$
ln\varrho_{q}(t,0)+\int_{0}^{\infty}\sigma(t-u, -u)(1+\varepsilon
u)\exp\{-\varepsilon u\}du=
$$
$$
\ln\varrho_{zub}(t)+ \int_{0}^{\infty}\sigma(t-u, -u)\varepsilon
uexp\{-\varepsilon u\}du;  \quad  \sigma(t-u,
-u)=dln\varrho_{q}(t-u, -u)/du,
$$
where $ln\varrho_{zub}(t)=ln\varrho_{q}(t,0)+\\
\int_{0}^{\infty}\sigma (t-u,-u)exp\{-\varepsilon
u\}du=\int_{0}^{\infty}\ln\varrho_{q}(t-u,-u)\varepsilon
\exp\{-\varepsilon u\}du$ is the Zubarev's form of the NSO,
$\sigma(t)= \partial\ln\varrho_{q}(t-u,-u)/\partial
u_{|u=0}=-\partial \ln q(t, 0)/\partial t$ is the entropy
production operator \cite{zub71}. It is seen from (\ref{exp2n})
that the logarithm of the NSO has an additional term in comparison
to the Zubarev's form. The source in the rhs of the Liouville
equation (or dissipative part of the Liouville operator
\cite{ra95}) equal $J=-\varepsilon
[ln\varrho(t)-ln\varrho_{zub}(t)]$, that is the system relaxes not
towards  $ln\varrho_{q}(t,0)$, like it is the case of Zubarev's
NSO, but towards $ln\varrho_{zub}(t)$. From the expression
(\ref{exp2n}) it is seen that introduced NSO contains amendments
to the Zubarev's NSO \cite{zub71,zub80}. The physical results
obtained with use of (\ref{exp2n}) also contains additional terms
in comparison to Zubarev's NSO. The additional terms describe the
influence of the lifetime finiteness on the kinetic processes. The
expressions for average fluxes \cite{zub71} averaged over
(\ref{exp2n}) have the form

\begin{equation}
<j^{m}(x)>= <j^{m}(x)>_{zub}+ \label{flu}
\end{equation}
$$
\sum_{n}\int\int_{-\infty}^{t}\varepsilon(t-t')exp\{\varepsilon
(t'-t)\}(j^{m}(x), j^{n}(x', t'-t))X_{m}(x', t`)dt'dx',
$$
where
$<j^{m}(x)>_{zub}=<j^{m}(x)>_{l}+\\
\sum_{n}\int\int_{-\infty}^{t} \exp\{\varepsilon
(t'-t)\}(j^{m}(x), j^{n}(x', t'-t)X_{m}(x', t')dt'dx'$ are fluxes
in the form obtained by Zubarev \cite{zub71}, $j^{n}$ are flux
operators, $X_{m}$ are corresponding thermodynamical forces;
$(j^{m}(x),j^{n}(x',t))=\beta^{-1}\int_{0}^{\beta}<j^{m}(x)(j^{n}(x',
t, i\tau)-\\
<j^{n}(x', t)>_{l})>_{l}d\tau$ are quantum time correlation
functions, $j^{n}(x', t, i\tau)=exp\{-\beta^{-1}A\tau\}j^{n}(x',
t)exp\{\beta^{-1}A\tau\}; A=\sum_{m}\int F_{m}(x,t)P_{m}(x)dx$.
The collision integrals of the generalized kinetic equation
\cite{zub71}, averaged over (\ref{exp2n}) have the amendments

\begin{equation}
S^{(2)}_{add}=-h^{-2}\int_{-\infty}^{0}dt\varepsilon
texp\{\varepsilon t\}<[H_{1}(t), [H_{1}, P_{k}]+ih\sum_{m}
P_{m}\frac{\partial S^{(1)}_{k}}{\partial <P_{m}>}]>_{q}
\label{kineq}
\end{equation}
to Zubarev result \cite{zub71}: $S^{(2)}=-h^{-2}\int_{-\infty}^{0}
dt \exp\{\varepsilon t\}<[H_{1}(t), [H_{1}, P_{k}]+\\
ih\sum_{m}P_{m}\frac{\partial S^{(1)}_{k}}{\partial
<P_{m}>}]>_{q}$, where the Hamiltonian of the system is
$H=H_{0}+H_{1}$, $H_{1}$ is the Hamiltonian of the interaction
which contains the longtime correlations \cite{ra95},
$S^{(1)}_{k}=\frac{<[P_{k}, H_{1}]>_{q}}{ih}$. The same is valid
for the generalized transport equations \cite{zub71}, kinetic
coefficients etc. Thus the selfdiffusion coefficient (or, to be
exact, its Laplace transform over time and space) obtained in
\cite{zub80} in the form

\begin{equation}
D(\omega , q)=\frac{q^{-2}\Phi(\omega , q)}{[1+ \frac{\Phi(\omega
, q)}{(i\omega -\varepsilon )}]}, \label{dif}
\end{equation}
where  $\Phi(\omega ,q)=\frac{\int_{0}^{\infty}
dtexp\{(i\omega-\varepsilon)t\}\int_{0}^{\beta}
<\dot{n}_{q}\dot{n}_{-q}(-t+ih\lambda)>d\lambda}{\int_{0}^{\beta}
<n_{q}n_{-q}(ih\lambda )>d\lambda}$, $n_{q}=\int
n(x)exp\{i(qx)\}dx=\sum_{j}\exp\{i(qx_{j})\}$, $n(x)=\delta
(x-x_{j})$, after use of (\ref{exp2n}) takes on the form

\begin{equation}
D(\omega, q)=\frac{q^{-2}[\Phi(\omega , q)+ \frac{\varepsilon
d\Phi(\omega , q)}{d(i\omega )}]}{1+\frac{\Phi (\omega ,
q)}{(i\omega -\varepsilon )}+\frac{ \varepsilon[ \frac{d\Phi
(\omega , q)}{d(i\omega)}-\frac{\Phi(\omega, q)}{(i\omega
-\varepsilon )}]}{(i\omega -\varepsilon )}}. \label{dif1}
\end{equation}

At   $\varepsilon\rightarrow0$, for infinitely large system in the
thermodynamic limit this expression (\ref{dif1}) coincides with
(\ref{dif}) at   $\varepsilon\rightarrow0$ \cite{zub80}. For
finite size systems (as well as for the case $\omega\rightarrow0$)
the results differ.

\section{Systems with infinite lifetime}
 \label{sect:3}

Above, as well as in work \cite{ry07}, additives to NSO in the
Zubarev's form are received for systems of the finite size, with
finite lifetime. We will show, as for systems with infinite
lifetime, for example, for systems of infinite volume, after
thermodynamic limiting transition, the same effects, which essence
in influence of the past of system, its histories, on its present
non-equilibrium state are fair.

In work \cite{ry07} it is shown, as changes in function $p_{q}(u)$
influences on non-equilibrium descriptions of the system. But for
those distributions $p_{q}(u)$, which are considered in
\cite{ry07} (gamma-distributions, (\ref{gam}), (\ref{exp2})) the
changes show up only for the systems of finite size with finite
lifetimes. Additions to unit in equation (\ref{sum}) becomes
vanishingly small to tend to infinity of sizes of the system and
its average lifetime, as in the model distribution (\ref{expDi})
used in Zubarev's NSO (\ref{NSO}). For the systems of finite size
and the exponential distribution results to nonzero additions in
expression (\ref{NSO}). Thus, these additions to NSO and proper
additions to kinetic equations, kinetic coefficients and other
non-equilibrium descriptions of the system, are an effect
finiteness of sizes and lifetime of the system, not choice of
distribution of lifetime of the system. We will find out, whether
there are distributions of lifetime of system for which and for
the infinitely large systems with infinitely large lifetime an
additional contribution to NSO differs from Zubarev'a NSO.

We will consider a few examples of task of function $p_{q}(u)$ in
(\ref{zu}), (\ref{NSO}). We will be limited to the examples of
task of the piecewise-continuous distributions, which result in
results different from works \cite{zub71,zub80}. There are
numerous experimental confirmations of such change of distribution
of lifetime of the system $p_{q}(u)$ on the time domain of life of
the system. This and transition to chaos and transition from the
laminar mode to turbulent one is also accompanied by the change of
distribution of $p_{q}(u)$. In works \cite{inoue,mantegna} is
shown transition of distribution of the first passage processes
from Gaussian regime to the not Levy conduct in some point of
time. Besides the real systems possess finite sizes and finite
lifetime. Therefore influence of surroundings on them is always
present. That a source not is equal to the zero and in a limit
infinitely large systems, is related to the openness of the
system, influence on her of surroundings

4a). We will put

\begin{equation}
p_{q}(u)= \big\{\left.
\begin{array}{l}
\frac{ka^{k}}{(u+a)^{k+1}}, \quad u<c,
\\
b\varepsilon\exp\{-\varepsilon u\}, \quad u\geq c;
\end{array}
\right. \label{pa}
\end{equation}

First part of distribution (\ref{pa}), Pareto distribution, in
work \cite{cox} got from the exponential distribution with the
random parameter of intensity $\rho$, when

$$
f_{T}(u)=\int_{0}^{\infty}\rho e^{-\rho u}f_{p}(\rho)d\rho
$$
at $f_{p}(\rho)=\frac{a(ka)^{k-1}exp\{-a\rho\}}{\Gamma(k)}$ is
gamma-distribution, $a=\frac{k}{\rho_{0}}$, $\rho_{0}$ is average
of gamma-distribution. From the condition of the normalization
$\int_{0}^{\infty}p_{q}(u)=1$ we will find a normalization
multiplier $b=e^{\varepsilon c}(\frac{a}{a+c})^{k}$. Average
lifetime $\langle\Gamma\rangle$ for distribution (\ref{pa}) is
equal

\begin{equation}
\langle\Gamma\rangle=\frac{a}{k-1}+(\frac{a}{a+c})^{k}[\frac{1}{\varepsilon}(1+\varepsilon
c)-\frac{(kc+a)}{(k-1)}]. \label{ava}
\end{equation}

The value (\ref{ava}) $\langle\Gamma\rangle \rightarrow \infty$ at
$\varepsilon\rightarrow0$. From (\ref{pa}) we find

\begin{equation}
-\int p_{q}(u)= \big\{\left.
\begin{array}{l}
(\frac{a}{u+a})^{k}, \quad u<c,
\\
b\exp\{-\varepsilon u\}, \quad u\geq c.
\end{array}
\right. \label{pa1}
\end{equation}

Source in right part of Liouville equation (\ref{li}) for
distribution (\ref{pa}) in accordance with expression (\ref{sour})
equal

\begin{equation}
J=\frac{k}{a}
\ln\varrho_{q}(t,0)-\int_{0}^{c}\frac{k(k+1)a^{k}}{(u+a)^{k+2}}Sdu-
e^{c\varepsilon}(\frac{a}{a+c})^{k}\int_{c}^{\infty}\varepsilon^{2}e^{-\varepsilon
u}Sdu, \label{soua}
\end{equation}
where $S=\ln\varrho_{q}(t-u,-u)$. Distribution of NSO (\ref{NSO})
in the case of distribution (\ref{pa}) equal

\begin{equation}
ln\varrho(t)= ln\varrho_{q}(t, 0)+\int_{0}^{\infty}e^{-\varepsilon
u}\frac{d \ln\varrho(t-u, -u)}{du}du+\Delta; \label{dea}
\end{equation}

$$
\Delta=\int_{0}^{c}[(\frac{a}{a+u})^{k}-e^{-\varepsilon u}]\sigma
du+\int_{c}^{\infty}[e^{\varepsilon
c}(\frac{a}{a+c})^{k}-1]e^{-\varepsilon u}\sigma du,
$$
where $\sigma=\frac{d \ln\varrho(t-u, -u)}{du};
ln_{zub}\varrho(t)=ln\varrho_{q}(t,
0)+\int_{0}^{\infty}e^{-\varepsilon u}\sigma du$ is distribution
got by Zubarev in \cite{zub71,zub80}, and $\Delta$ are amendments
to him.

4b). We will consider distribution of kind now
\begin{equation}
p_{q}(u)= \big\{\left.
\begin{array}{l}
d, \quad u<c,
\\
\varepsilon\exp\{-\varepsilon u\}, \quad u\geq c;
\end{array}
\right. \label{pqb}
\end{equation}

From the condition of the normalization we find a value
$d=\frac{1}{c}(1-e^{-\varepsilon c})$. Average lifetime
\begin{equation}
\langle\Gamma\rangle=\frac{dc^{2}}{2}+\frac{1}{\varepsilon}e^{-\varepsilon
c}(1+\varepsilon c).\label{avb}
\end{equation}
Average lifetime $\langle\Gamma\rangle\rightarrow\infty$ at
$\varepsilon\rightarrow0$. Source (\ref{sour}) in equation
(\ref{li}) in this case equal
\begin{equation}
J=d\ln\varrho_{q}(t,0)-\int_{c}^{\infty}\varepsilon^{2}e^{-\varepsilon
u}Sdu. \label{soub }
\end{equation}

NSO is equal
\begin{equation}
ln\varrho(t)= ln\varrho_{zub}(t, 0)-\int_{0}^{c}[e^{-\varepsilon
u}+\frac{1}{c}(1-e^{-\varepsilon c})u]\sigma du; \label{deb}
\end{equation}
$$
\Delta=-\int_{0}^{c}[e^{-\varepsilon
u}+\frac{1}{c}(1-e^{-\varepsilon c})u]\sigma du.
$$

We get in this case, that in an additional item memory of the
system is limited by a size, is observed effect of limited memory.
It is possible to consider and other examples of task
distributions $p_{q}(u)$ which reduce to limited memory.

4c). For the exponential density of distribution with different
intensities in different temporal intervals
\begin{equation}
p_{q}(u)= \big\{\left.
\begin{array}{l}
\varepsilon_{1}\exp\{-\varepsilon_{1} u\}, \quad u<c,
\\
b\varepsilon_{2}\exp\{-\varepsilon_{2} u\}, \quad u\geq c,
\end{array}
\right. \label{pqc}
\end{equation}
from the condition of the normalization finds that
$b=e^{c(\varepsilon_{2}-\varepsilon_{1})}$;

\begin{equation}
\langle\Gamma\rangle=\frac{1}{\varepsilon_{1}}[1-e^{-\varepsilon_{1}
c}(1+\varepsilon_{1}
c)]+\frac{1}{\varepsilon_{2}}e^{-\varepsilon_{1}
c}(1+\varepsilon_{2} c). \label{avc}
\end{equation}

At $\langle\Gamma\rangle\rightarrow\infty$,
$\varepsilon_{2}\rightarrow 0$.

$$
J=\varepsilon_{1}\ln\varrho_{q}(t,0)-\int_{0}^{c}\varepsilon_{1}^{2}e^{-\varepsilon_{1}
u}Sdu-e^{c(\varepsilon_{2}-\varepsilon_{1})}\int_{c}^{\infty}\varepsilon_{2}^{2}e^{-\varepsilon_{2}
u}Sdu;
$$
\begin{equation}
ln\varrho(t)= ln\varrho_{zub}(t)+\Delta; \label{dec}
\end{equation}
$$
\Delta=\int_{0}^{c}[e^{-\varepsilon_{1} u}-e^{-\varepsilon_{2}
u}]\sigma
du+\int_{c}^{\infty}[e^{c(\varepsilon_{2}-\varepsilon_{1})}-1]e^{-\varepsilon_{2}
u}\sigma du;
$$
$$
\Delta_{\varepsilon_{2}\rightarrow 0}\rightarrow
\int_{0}^{\infty}[e^{-\varepsilon_{1} u}-1] \sigma du.
$$

4d). We will consider yet distribution of kind
\begin{equation}
p_{q}(u)= \big\{\left.
\begin{array}{l}
\varepsilon^{2}u\exp\{-\varepsilon u\}, \quad u<c,
\\
b\frac{\exp\{-\gamma u\}}{[1+\frac{q-1}{q}a\gamma \exp\{-\gamma
au\}]^{\frac{1}{(q-1)}}}, \quad u\geq c.
\end{array}
\right. \label{pqd}
\end{equation}
The second part of distribution (\ref{pqd})can be received from
results of works \cite{ryaz} and like to Tsallis distributions
\cite{tsal}. For (\ref{pqd}) $b=\frac{\gamma
a[\frac{(q-1)}{q}\gamma a]^{\frac{1}{a}}[1-\exp\{-\varepsilon
c\}(1+\exp\{-\varepsilon
c\})]}{B_{(1-p,1)}(1-\frac{1}{(q-1)},\frac{1}{a})}$, where
$B(1-p,1)$ is incomplete beta function \cite{abr},
$p=\frac{(q-1)}{q}a\gamma\exp\{-\gamma ac\}$.

Average lifetime
$$
\langle\Gamma\rangle=\frac{2}{\varepsilon}[1-e^{-\varepsilon
c}(1+\varepsilon c+(\varepsilon c)^{2}/2)]+
$$
$$
+e^{-\varepsilon c}(1+\varepsilon
c)\frac{\Gamma^{2}(\frac{1}{a})_{3}F_{2}(\frac{1}{a},\frac{1}{a},\frac{1}{(q-1)};1+\frac{1}{a},1+\frac{1}{a};p)}
{(a^{2}\gamma)_{2}F_{1}(\frac{1}{a},\frac{1}{(q-1)};1+\frac{1}{a};p)},
$$
where $\Gamma(.)$ is gamma function, $_{n}F_{m}$ is
hypergeometrical functions \cite{abr}, tends to infinity at
$\gamma\rightarrow0$. In this case value of the normalization
$b\rightarrow0$. Therefore at
$\langle\Gamma\rangle\rightarrow\infty, \gamma\rightarrow0$ and
$b\rightarrow0$;

$$
ln\varrho(t)= ln\varrho_{q}(t,0)+\int_{0}^{c}e^{-\varepsilon
u}(1+\varepsilon u)\sigma du; J=-\int_{0}^{c}e^{-\varepsilon
u}\varepsilon^{2}(1-\varepsilon u)Sdu.
$$
The effect of limited memory shows up in this case obviously.

Why the examples of this section differ from examples of section
3? In interpretation \cite{zub80} fluctuate the random value
$t_{0}$ in $u=t-t_{0}$. In \cite{zub80} limiting transition is
conducted for the parameter $\varepsilon, \varepsilon\rightarrow
0$ in the exponential distribution
$p_{q}(u)=\varepsilon\exp\{-\varepsilon u\}$ after thermodynamical
limiting transition. In interpretation of work \cite{ry01} it
corresponds to that mean lifetime of the system
$\langle\Gamma\rangle=\langle
t-t_{0}\rangle=1/\varepsilon\rightarrow\infty$. But middle
intervals between random shoves infinitely increase, exceeding
lifetime of the system. Therefore an item with a source in
Liouville equation applies in a zero. If there is the change of
distribution $p_{q}(u)$ on the time domain of life of the system,
as in examples 4a)-4d), influence of surroundings, followed this
change with, remains in the time domain of life even at tendency
to infinity of mean lifetime.

\section{Application of distributions of section 4 to research of conductivity}
 \label{sect:4}

On the example of determination of conductivity we will consider,
in what consequences the change of type of functions $p_{q}(u)$
and $\varrho(t)$ results as compared to an exponential law for
$p_{q}(u)$, used in work \cite{zub96}.

Determination of conductivity by a method NSO is considered in
works \cite{kal74,rep81,rep91a,bobr}. In this section we will
consider the transport of charges in the electric field, as linear
reaction on mechanical perturbation, conductivity in the linear
approaching, following results of work \cite{zub96} and, as in
\cite{zub96}, we will be limited to the important special case -
reaction of the equilibrium system on the spatially homogeneous
variable field.
$$
\vec{E}^{0}(t)=\int_{-\infty}^{\infty}\frac{d\omega}{2\pi}e^{-i\omega
t}\tilde{\vec{E}}^{0}(\omega).
$$
Hamiltonian of perturbation is given by a formula
$$
H^{1}_{t}=-\vec{P}\vec{E}^{0}(t),
$$
where $\vec{P}$ is operator proper to the vector of polarization
of the system. In coordinate representation this operator is
written down as
$$
\vec{P}=\sum_{i}e_{i}\vec{r}_{i},
$$
where $e_{i}$ is charge of particle, and $\vec{r}_{i}$ is its
radius vector. The operator of current is equal
$$
\vec{J}=\dot{\vec{P}}=e\sum_{j}\dot{\vec{r}}_{j}=\frac{e}{m}\sum_{j}\vec{p}_{j},
$$
where $\vec{p}_{j}$ is particle momentum, $m$ is mass of a
particle. We will choose a model in which coulomb interaction is
taken into account as self-consistent screening the field, i.e. we
will consider that $\vec{E}=\vec{E}^{0}$. Most essential
difference from work \cite{zub96} at consideration of this problem
consists of replacement of Laplace transformation used in
\cite{zub96}, form
\begin{equation}
\langle A;B
\rangle_{\omega+i\varepsilon}=\int_{0}^{\infty}dte^{i(\omega+i\varepsilon)t}(A(t),B),
(\varepsilon>0), \label{la}
\end{equation}
where $(A(t),B(t'))=\int_{0}^{1}dxTr\{\Delta A(t)\Delta
B(t'+i\beta \hbar x)\varrho_{eq}\}$ is temporal correlation
function \cite{zub96}, by other integral transformation. So, for
an example 4a) with distribution of form (\ref{pa}), (\ref{pa1})
expression (\ref{la}) it is replaced on
\begin{equation}
\langle A;B \rangle_{\omega;a,k}+e^{\varepsilon
c}(\frac{a}{a+c})^{k}\langle A;B
\rangle_{\omega+i\varepsilon;(c,\infty)}, \label{la1}
\end{equation}
where
\begin{equation}
\langle A;B \rangle_{\omega;a,k}=\int_{0}^{c}dte^{i\omega
t}(\frac{a}{a+t})^{k}(A(t),B);           \langle A;B
\rangle_{\omega+i\varepsilon;(c,\infty)}=\int_{c}^{\infty}dte^{i(\omega+i\varepsilon)t}(A(t),B).
\label{la2}
\end{equation}

We will consider a isotropic environment in which tensor of
conductivity is diagonal. In work \cite{zub96} for Laplace
transformation of kind (\ref{la}) expression is got for specific
resistance $\rho(\omega)$ of kind
\begin{equation}
\rho(\omega)=\frac{1}{\sigma(\omega)}=\frac{3V}{\beta(\vec{J},\vec{J})}[-i\omega+M];
\label{re1}
\end{equation}

\begin{equation}
M=\frac{\langle\dot{\vec{J}};\dot{\vec{J}}\rangle_{\omega+i\varepsilon}}
{(\vec{J},\vec{J})+\langle\dot{\vec{J}};\vec{J}\rangle_{\omega+i\varepsilon}},
\label{m}
\end{equation}
where $V$ is the volume of the system, $\beta$ is inverse
temperature, $\sigma(\omega)$ is scalar coefficient of
conductivity or simply conductivity. In the examples considered
below in expression (\ref{m}) a value $M$ will change. Doing the
operations conducted in work \cite{zub96}, with replacement of
expression (\ref{la}) by expression (\ref{la1}), we will get for
this case in place of correlation (\ref{m}) more difficult
expression of kind
$$
M=\frac{\langle\dot{\vec{J}};\dot{\vec{J}}\rangle_{\omega;a,k}+\frac{i\omega}{i(\omega+i\varepsilon)}e^{i\omega
c}(\frac{a}{a+c})^{k}\langle\dot{\vec{J}};\dot{\vec{J}}\rangle_{\omega+i\varepsilon;(c,\infty)}}
{K};
$$
$$
K=(\vec{J}(0),\vec{J})-\frac{k}{a}\langle\vec{J};\vec{J}\rangle_{\omega;a,k+1}-(1-\frac{i\omega}{i(\omega+i\varepsilon)})
e^{i\omega c}(\frac{a}{a+c})^{k}(\vec{J}(c),\vec{J})+
$$
$$
+\langle\dot{\vec{J}};\vec{J}\rangle_{\omega;a,k}+\frac{i\omega}{i(\omega+i\varepsilon)}e^{i\varepsilon
c}
\langle\dot{\vec{J}};\vec{J}\rangle_{\omega+i\varepsilon;(c,\infty)}.
$$
At $\varepsilon\rightarrow0$
$$
M=\frac{\langle\dot{\vec{J}};\dot{\vec{J}}\rangle_{\omega;a,k}+e^{i\omega
c}(\frac{a}{a+c})^{k}\langle\dot{\vec{J}};\dot{\vec{J}}\rangle_{\omega;(c,\infty)}}
{(\vec{J}(0),\vec{J})-\frac{k}{a}\langle\vec{J};\vec{J}\rangle_{\omega;a,k+1}+
\langle\dot{\vec{J}};\vec{J}\rangle_{\omega;a,k}+
\langle\dot{\vec{J}};\vec{J}\rangle_{\omega;(c,\infty)}}.
$$
For distribution (\ref{pqb}) in 4b) Laplace transformation of kind
(\ref{la}) it is replaced on
$$
-d\langle A;B \rangle_{\omega;c,t}+\langle A;B
\rangle_{\omega+i\varepsilon;(c,\infty)},
$$
where $\langle A;B \rangle_{\omega;c,t}=\int_{0}^{c}dte^{i\omega
t}t(A(t),B)$, the value $d$ is given in (\ref{pqb}), (\ref{avb}),
$\langle A;B \rangle_{\omega+i\varepsilon;(c,\infty)}$ is given in
(\ref{la2}). In place of expression (\ref{m}) in this case we will
get expression
$$
M=\frac{\frac{i\omega}{i(\omega+i\varepsilon)}\langle\dot{\vec{J}};
\dot{\vec{J}}\rangle_{\omega+i\varepsilon;(c,\infty)}-d[\langle\dot{\vec{J}};\dot{\vec{J}}\rangle_{\omega;c,t}+
\langle\vec{J};\dot{\vec{J}}\rangle_{\omega;c,t=1}]} {K_{1}},
$$
$$
K_{1}=\frac{i\omega}{i(\omega+i\varepsilon)}\langle\dot{\vec{J}};\vec{J}\rangle_{\omega+i\varepsilon;(c,\infty)}-
d[\langle\dot{\vec{J}};\vec{J}\rangle_{\omega;c,t}+
$$
$$
+\langle\vec{J};\vec{J}\rangle_{\omega;c,t=1}]+(dc+\frac{i\omega}{i(\omega+i\varepsilon)}e^{-\varepsilon
c}) e^{i\omega c}(\vec{J}(c);\vec{J}).
$$
If $\varepsilon\rightarrow0, \langle \Gamma
\rangle\rightarrow\infty$, $d\rightarrow0$, and
$$
M=\frac{\langle\dot{\vec{J}};\dot{\vec{J}}\rangle_{\omega;(c,\infty)}}
{e^{i\omega
c}(\vec{J}(c),\vec{J})+\langle\dot{\vec{J}};\vec{J}\rangle_{\omega;(c,\infty)}}.
$$
This expression at the small values $c$ low differs from
(\ref{m}). For a case 4c) with distribution $p_{q}(u)$ of kind
(\ref{pqc}) Laplace transformation (\ref{la}) it is substituted by
a value
$$
\langle A;B
\rangle_{\omega+i\varepsilon;(0,c)}+e^{c(\varepsilon_{2}-\varepsilon_{1})}\langle
A;B \rangle_{\omega+i\varepsilon_{2};(c,\infty)},
$$
where $\langle
A;B\rangle_{\omega+i\varepsilon;(0,c)}=\int_{0}^{c}dte^{i(\omega+i\varepsilon)
t}(A(t),B)$, $\langle A;B
\rangle_{\omega+i\varepsilon_{2};(c,\infty)}$ is given in
(\ref{la2}) and value $M$ from (\ref{m}) it is substituted by a
value
$$
M=\frac{\langle\dot{\vec{J}};\dot{\vec{J}}\rangle_{\omega+i\varepsilon_{1};(0,c)}+
e^{c(\varepsilon_{2}-\varepsilon_{1})}\frac{i(\omega+i\varepsilon_{1})}{i(\omega+i\varepsilon_{2})}
\langle\dot{\vec{J}};\dot{\vec{J}}\rangle_{\omega+i\varepsilon_{2};(c,\infty)}}{K_{2}},
$$
$$
K_{2}=(\vec{J}(0),\vec{J})-e^{i(\omega+i\varepsilon_{1})c}(\vec{J}(c);\vec{J})+
\langle\dot{\vec{J}};\vec{J}\rangle_{\omega+i\varepsilon_{1};(0,c)}+
$$
$$
+e^{c(\varepsilon_{2}-\varepsilon_{1})}\frac{i(\omega+i\varepsilon_{1})}{i(\omega+i\varepsilon_{2})}
[e^{i(\omega+i\varepsilon_{2})c}(\vec{J}(c),\vec{J})+
\langle\dot{\vec{J}};\vec{J}\rangle_{\omega+i\varepsilon_{2};(c,\infty)}].
$$
At $\langle \Gamma \rangle\rightarrow\infty$ and
$\varepsilon_{2}\rightarrow0$ the value $M$ changes unessential,
assuming a form
$$
M=\frac{\langle\dot{\vec{J}};\dot{\vec{J}}\rangle_{\omega+i\varepsilon_{1};(0,c)}+
e^{-c\varepsilon_{1}}\frac{i(\omega+i\varepsilon_{1})}{i\omega}
\langle\dot{\vec{J}};\dot{\vec{J}}\rangle_{\omega;(c,\infty)}}{K_{3}}
$$
$$
K_{3}=(\vec{J}(0),\vec{J})-e^{i(\omega+i\varepsilon_{1})c}(\vec{J}(c),\vec{J})+
\langle\dot{\vec{J}};\vec{J}\rangle_{\omega+i\varepsilon_{1};(0,c)}+
$$
$$
+e^{-c\varepsilon_{1}}\frac{i(\omega+i\varepsilon_{1})}{i\omega}
[e^{i\omega
c}(\vec{J}(c),\vec{J})+\langle\dot{\vec{J}};\vec{J}\rangle_{\omega;(c,\infty)}].
$$

At the small values of value $\varepsilon_{1}$ this expression
close to (\ref{m}). From (\ref{avc}) it is visible that
$\lim_{\varepsilon_{2}\rightarrow0}\varepsilon_{2}\langle \Gamma
\rangle=e^{-\varepsilon_{1}c}$.

Unfortunately, to compare the got results to the experiment
difficultly, because parameters $c,a,k,\varepsilon_{1}$ are
unknown. But, apparently, through the choice of type of
distribution $p_{q}(u)$ and selection of his parameters it is
possible to obtain very good accordance with experimental results.

\section{Comparison with the theory of  transport processes by McLennan}
 \label{sect:5}

In works \cite{mc1,mc2} the statistical theory of  transport
processes, based on introduction of external forces of unpotential
behavior, describing influence of surroundings on this system, is
built. In the Appendix II to work \cite{zub71} this theory is
compared with the method of NSO.

The function of distribution of the complete system $f_{u}$ is
described by Liouville equation
\begin{equation}
\frac{\partial f_{u}}{\partial t}+\{f_{u},H_{u}\}=0, \label{li1}
\end{equation}
where $\{…,…\}$ are classic Poisson brackets, $H_{u}=H+H_{s}+U$,
$H$ is the Hamiltonian of the concerned system, $H_{s}$ is
Hamiltonian of surroundings, $U$ is Hamiltonian of interaction of
the system with surroundings.  The functions of distributions of
the concerned system $f$ and surroundings $g$ are accordingly
equal
$$
f=\int f_{u}d\Gamma_{s},     g=\int f_{u}d\Gamma,
$$
where $d\Gamma_{s}$ and $d\Gamma$ are elements of phase volume of
surroundings and concerned system. Integrating (\ref{li1}) on
phase space of surroundings $d\Gamma_{s}$, we get equation of
motion for $f$, containing a source. At introduction of function
$X$ describing correlation of the system with surroundings
\begin{equation}
f_{u}=fgX, \label{x1}
\end{equation}
equation for $f$ is written down in a form
\begin{equation}
\frac{\partial f}{\partial t}+\{f,H\}+\frac{\partial
(fF_{\alpha})}{\partial p_{\alpha}}=0, \label{li2}
\end{equation}
where
\begin{equation}
F_{\alpha}=-\int gX\frac{\partial U}{\partial
q_{\alpha}}d\Gamma_{s}, \label{x1}
\end{equation}
$q_{\alpha}$ and $p_{\alpha}$ are coordinates and impulses of the
system, in (\ref{li2}) summation up is assumed on $\alpha$. A
value $F_{\alpha}$ makes sense "force" presenting the action of
surroundings on the system. In \cite{mc1,mc2} from the physical
considering justice of expression is assumed
\begin{equation}
\frac{\partial F_{\alpha}}{\partial
p_{\alpha}}=-\int\vec{j}_{s}(\vec{x})d\vec{s}, \label{for}
\end{equation}
where $\vec{j}_{s}(\vec{x}$ is density of entropy flow (including
the work accomplished above the system), $d\vec{s}$ is element of
surface limiting the system. For the function of distribution in
\cite{zub71} expression coincident with expression for NSO at
$\varepsilon\rightarrow0$ is got. If to accept, that correlation
between the system and surroundings results in replacement of
expression (\ref{for}) on expression
$$
\frac{\partial F_{\alpha}}{\partial
p_{\alpha}}=-\int\vec{j}_{s}(\vec{x})Y(\vec{x},t)d\vec{s},
$$
and $Y(x,t)=-\int p_{q}(t)dt$, after integration on time,
conducted at the decision of equation for the function of
distribution, we will get for the function of distribution of the
system expressions coincident with expressions of the this work.

Indeed, the function of distribution of time of past life
$t-t_{0}$ of the system $p_{q}(u)$ depends and from properties of
the system and from surroundings, as $X$. A function $X$ must
depend on time. Then, for example, for expression 4a) with
distribution (\ref{pa}), (\ref{pa1}) function $-\int p_{q}(t)dt$
related to $X(t)$, to the moment of time $c$ decreases, as
$(\frac{a}{a+t})^{k}$, and from a moment $c$ assumes a form
$(\frac{a}{a+c})^{k}e^{-\varepsilon c}e^{-\varepsilon t}$,
$\varepsilon\rightarrow0$ after thermodynamic limiting transition
and tendency $\langle\Gamma\rangle\rightarrow\infty$. Time $c$
comes forward in a role of time of establishment of stationary
flows. For an example 4c) with distribution (\ref{pqc}) a function
$Y(x,t)$ decreases as $e^{-\varepsilon_{1}t}$ at $t<c$ and as
$e^{c(\varepsilon_{1}-\varepsilon_{2})}e^{-\varepsilon_{2}t}$ at
$t\geq c$, $\varepsilon_{2}\rightarrow0$ at
$\langle\Gamma\rangle\rightarrow\infty$. Thus, the function of
distribution $p_{q}(u)$ of time of past life of the system (more
precisely, integral from her with a reverse sign) can be
interpreted and in connection with the function of correlation $X$
between the system and surroundings.

\section{The conclusion}
 \label{sect:4}

As it is specified in work \cite{rau}, existence of time scales
and a stream of the information from slow degrees of freedom to
fast create irreversibility of the macroscopical description. The
information continuously passes from slow to fast degrees of
freedom. This stream of the information leads to irreversibility.
The information thus is not lost, and passes in the form
inaccessible to research on Markovian level of the description.
For example, for the rarefied gas the information is transferred
from one-partial observables to multipartial correlations. In work
\cite{ry01} values  $\varepsilon=1/\langle\Gamma\rangle$ and
$p_{q}(u)=\varepsilon\exp\{-\varepsilon u\}$ are expressed through
the operator of entropy production and, according to results
\cite{rau}, - through a stream of the information from relevant to
irrelevant degrees of freedom. Introduction in NSO function
$p_{q}(u)$ corresponds to specification of the description by
means of the effective account of communication with irrelevant
degrees of freedom. In the present work it is shown, how it is
possible to spend specification the description of effects of
memory within the limits of method NSO, more detailed account of
influence on evolution of system of quickly varying variables
through the specified and expanded kind of density of function of
distribution of time the lived system of a life.

In many physical problems finiteness of lifetime can be neglected.
Then $\varepsilon\sim1/\langle\Gamma\rangle\rightarrow0$. For
example, for a case of evaporation of drops of a liquid it is
possible to show \cite{ryDr}, that non-equilibrium characteristics
depend from $\exp\{y^{2}\}; y=\varepsilon/(2\lambda_{2})^{1/2},
\lambda_{2}$ is the second moment of correlation function of the
fluxes averaged on quasi-equilibrium distribution. Estimations
show, what even at the minimum values of lifetime of drops
(generally - finite size) and the maximum values size
$y=\varepsilon/(2\lambda_{2})^{1/2}\leq 10^{-5}$. Therefore
finiteness of values $\langle\Gamma\rangle$ and $\varepsilon$ does
not influence on behaviour of system and it is possible to
consider $\varepsilon=0$. However in some situations it is
necessary to consider finiteness of lifetime
$\langle\Gamma\rangle$ and values  $\varepsilon> 0$. For example,
for nanodrops already it is necessary to consider effect of
finiteness of their lifetime. For lifetime of neutrons in a
nuclear reactor in work \cite{ry01} the equation for
$\varepsilon=1/\langle\Gamma\rangle$ which decision leads to
expression for average lifetime of neutrons which coincides with
the so-called period of a reactor is received. In work \cite{ryAt}
account of finiteness of lifetime of neutrons result to correct
distribution of neutrons energy.

Use of distributions (\ref{gam}), (\ref{exp2}) and several more
obvious forms of lifetime distribution in quality $p_{q}(u)$ leads
to a conclusion, that the deviation received by means of these
distributions values $\ln\varrho(t)$ from $\ln\varrho_{zub}(t)$ is
no more $\varepsilon\sim1/\langle\Gamma\rangle$. Therefore in
expression (\ref{NSO}) additives to the result received by
Zubarev, are proportional $\varepsilon$. This result corresponds
to mathematical results of the theory asymptotical phase
integration of complex systems \cite{tur} according to which
distribution of lifetime looks like $p_{q}(u)=\exp\{-\varepsilon
u\}+\lambda\varphi_{1}(u)+\lambda^{2}\varphi_{2}(u)+...$ , where
the parameter of smallness $\lambda$ in our case corresponds to
value $\varepsilon\sim1/\langle\Gamma\rangle$. Generally the
parameter $\lambda$ can be any.

For distributions of kind (\ref{pa}), (\ref{pqb}), (\ref{pqc}),
having a various form for different times, additives to Zubarev's
NSO are distinct from zero and for infinitely large systems with
infinitely large lifetimes. In the present work it is shown, that
it is possible, for example, for distributions of lifetime of the
system, having a various appearance at different stages of
evolution of system. Such behaviour will be coordinated with known
division of process of evolution of system into a number of stages
\cite{bog}. For some distributions the effect of "finite memory"
when only the limited interval of the past influences on behaviour
of system is observed.

Probably, similar results will appear useful, for example, in
researches of small systems. All greater value is acquired by
importance of description of the systems in mesoscopical scales. A
number of the results following from interpretation of NSO and
$p_{q}(u)$ as density of lifetime distribution of system
\cite{ry01}, it is possible to receive from the stochastic theory
of storage \cite{prab} and theories of queues. For example, in
\cite{prab} the general result that the random variable of the
period of employment (lifetime) has absolutely continuous
distribution $p_{q}(u)=g(u,x)=xk(u-x,u), u>x>0$ is resulted;
$g(u,x)=0$ in other cases, where $k(x,t)$ is absolutely continuous
distribution for value $X(t)$ - input to system.

The form of distribution chosen by Zubarev for lifetime represents
limiting distribution. The choice of lifetime distribution in NSO
is connected with the account of influence of the past of system,
its physical features, for the present moment, for example, with
the account only age of system, as in Zubarev's NSO
\cite{zub71,zub80,ry01} at $\varepsilon>0$, or with more detailed
characteristic of the past evolution of system. The received
results are essential in cases when it is impossible to neglect
effects of memory when memory time there is not little. The
analysis of time scales as it is noted in \cite{rau}, it is
necessary to spend in each problem.

To determination of type of distribution $p_{q}(u)$ it is possible
to apply general principles, for example, principle of maximum of
entropy, as this it was done in the work \cite{plas} for
determination of type of source in kinetic equation. It was marked
that exists many possibilities of choice of function $p_{q}(u)$.
But by virtue of a number of the reasons not all functions can be
used in NSO. So, in work \cite{der85} it was marked that NSO can
exist only if corresponding retarded theories are memory
renormalizable. In general case offered approach can be
interpreted as working out in detail of history of evolution of
the system, clarification of different stages of its conduct.


\begin{thebibliography}{99}

\bibitem{zub71} D.N. Zubarev, \emph{Nonequilibrium statistical
thermodynamics}, Plenum-Consultants Bureau, New York (1974).

\bibitem{zub80} D.N. Zubarev, in Reviews of Science and Technology: \emph{Modern Problems of
Mathematics}, \textbf{15}, (1980), 131-226, (in Russian) ed. by R.
B. Gamkrelidze, Nauka, Moscow [English Transl.: J. Soviet Math.
\textbf{16} (1981), 1509].

\bibitem{kal} D.N. Zubarev, V.P. Kalashnikov, A time-irreversible generalized master equation,
\emph{Phys. Lett. A}, \textbf{34A}, (1971), 311-312.

\bibitem{kal71} D.N. Zubarev, V.P. Kalashnikov, The derivation of time-irreversible generalized master
equation, \emph{Physica}, \textbf{56}, (1971), 345-364.

\bibitem{zub96} D.N.Zubarev, V.Morozov, G.R\"{o}pke, \emph{Statistical mechanics of nonequilibrium Processes.
Vol. 1, Basic Concepts, Kinetic Theory}, Akad. Verl., Berlin,
(1996).

\bibitem{ry01} V.V.Ryazanov, Lifetime of System and Nonequilibrium Statistical Operator Method,
\emph{Fortschritte der Phusik/Progress of Physics}, \textbf{49}
(2001), 885-893.

\bibitem{ra95} R. Luzzi, A.R. Vasconcellos and J.G. Ramos, \emph{Statistical Foundations of Irreversible Thermodynamics},
Teubner-BertelsmannSpringer, Stutgart, Germany (2000).

\bibitem{ra99} J.G. Ramos, A.R. Vasconcellos and R. Luzzi, On the thermodynamics
of far-from-equilibrium dissipative systems, \emph{Fortschr.
Phys./Progr. Phys.}, \textbf{47} (1999), 937-954.

\bibitem{str61} R.L. Stratonovich, \emph{The elected questions of the fluctuations
theory in a radio engineering}, Gordon and Breach, New York
(1967).

\bibitem{ry07} V.V. Ryazanov, Nonequilibrium statistical
operator for systems with finite lifetime, \emph{Low Temperature
Physics}, \textbf{33 } (2007), 1049-1053.

\bibitem{kirk} J.G. Kirkwood, The statistical mechanical
theory of transport processes. I.General theory,
\emph{J.Chem.Phys.}, \textbf{14} (1946), 180-192; II. Transport in
gases, \emph{J.Chem.Phys.}, \textbf{15} (1946), 72-84.

\bibitem{Prig} I. Prigogine, \emph{From Being to Becoming}, Freeman, San Francisco (1980).

\bibitem{rep83} R. Der, G.R\"{o}pke, Influence of infinitesimal source
terms in the Liouville equation (zubarev's method) on macroscopic
evolution equations, \emph{Phys. Lett.A}, \textbf{95A}, (1983),
347-349.

\bibitem{der85} R. Der, On the retarded solution of the Liouville equation
and the definition of entropy in kinetic theory, \emph{PhysicaA},
\textbf{132A}, (1985), 74-93.

\bibitem{rau} J.Rau, B.Muller, From reversible quantummicrodynamics to irreversible quantum
transport, \emph{Physics Reports}, \textbf{272} (1996), 1-59.

\bibitem{bog} N.N. Bogoliubov, \emph{The Problem of Dynamical Theory in Statistical Physics} (in Russian),
Izdatel'stvo O.G.I.Z. Gostekhizdat, Moscow-Leningrad (1946); N.N.
Bogoliubov, in \emph{Studies in Statistical Mechanics} I, edited
by J. de Boer and G.E. Uhlenbeck, North Holland, Amsterdam (1962),
pp. 4-118.

\bibitem{mor} V.G.Morozov, G.R\"{o}pke, Zubarev's method of a nonequilibrium statistical operator
and some challenges in the theory of irreversible processes,
\emph{Condensed Matter Physics}, \textbf{1} (1998), 673-686.

\bibitem{fe} W. Feller, \emph{An Introduction to Probability Theory and its Applications},
vol.2, J.Wiley, New York (1971).


\bibitem{klim} Yu.L. Klimontovich, \emph{Statistical Theory of Open Systems}, Kluwer Acad.
 Publ., Dordrecht (1995).

\bibitem{gr} M.S. Green, Markoff Random Processes and the Statistical Mechanics of Time-Dependent Phenomena,
\emph{J. Chem. Phys.} \textbf{20} (1952), 1281-1297 ; ibid.
\textbf{22} (1954), 398-409.


\bibitem{mo} H. Mori, I. Oppenheim and J. Ross, in \emph{Studies in Statistical Mechanics} I, edited by
J.de Boer and G.E.Uhlenbeck, North-Holland, Amsterdam, (1962), pp. 217-298


\bibitem{co} D.R. Cox, \emph{Renewal theory}, John Wiley, London: Methuen; New York (1961).

\bibitem{cox} D.R. Cox and D. Oakes, \emph{Analysis of Survival Data}, Chapman and Hall, London,
New York (1984).


\bibitem{tur} V.S. Korolyuk and A.F. Turbin, \emph{Mathematical Foundations of the State Lumping of Large Systems}
Kluwer Acad.Publ., Dordrecht, Boston/London (1993).

\bibitem{str95} R.L. Stratonovich, To the cleanly dynamic theory of spontaneous
disintegration of complex molecules, \emph{Soviet Physics JETP},
\textbf{109} (1995), 1328-1336.

\bibitem{str96} R.L. Stratonovich, O.A. Chichigina, Calculation of permanent
spontaneous disintegration of clusters from identical atoms on a
dynamic theory, \emph{Soviet Physics JETP}, \textbf{110 }(1996),
1284-1291.

\bibitem{gasp93} P. Gaspard, What is the role of
chaotic scattering in irreversible processes, \emph{Chaos},
\textbf{3} (1993), 427-442.

\bibitem{gasp95} P. Gaspard, J.R. Dorfman, Chaotic Scattering theory,
thermodynamic formalism, and transport coefficients,
\emph{Phys.Rev.E}. \textbf{52} (1995), 3525-3552.

\bibitem{dorf} J.R. Dorfman,  P. Gaspard,
Chaotic scattering theory of transport and reaction-rate
coefficients , \emph{Phys.Rev.E.}, \textbf{51} (1995), 28-33.

\bibitem{str66} R.L. Stratonovich, \emph{Theory of information}, Sovetskoe radio, Moskow, (1966)
(In Russian).

\bibitem{dew} R. Dewar, Information theory explanation of the fluctuation
theorem, maximum entropy production and self-organized criticality
in non-equilibrium stationary states, \emph{J.Phys.A: Math. Gen}.
\textbf{36} (2003), 631-641.

\bibitem{mart} G.A. Martynov, Thermodynamics and Hydrodynamics (Statistical Foundations): 1. The Problem
Setup, \emph{Theoretical and Mathematical Physics},  \textbf{133}
(2002), 1421-1429.

\bibitem{prab} N.U. Prabhu, \emph{Stochastic Storage Processes}, Springer, Berlin, (1980).

\bibitem{inoue} Jun-ichi Inoue and Naoya Sazuka, Crossover between Le'vy and
Gaussian regimes in first-passage processes, \emph{Physical Review
E}, PP 021111(9) August 2007; Volume \textbf{76}, Number 2

\bibitem{mantegna} R.N. Mantegna and H.E. Stanley, \emph{Phys. Rev. Lett}, \textbf{73}, (1994) 2946.

\bibitem{ryaz} V.V. Ryazanov, First-Passage Time: A Conception Leading to Superstatistics.
I. Superstatistics with Discrete Distributions,
arXiv:physics/0509098; V.V. Ryazanov, First-Passage Time: A
Conception Leading to Superstatistics. II. Continuous
Distributions and their Applications,  arXiv:physics/0509099; V.V.
Ryazanov, S.G. Shpyrko, First-passage time: a conception leading
to superstatistics, \emph{Condensed Matter Physics}, \textbf{9},
(2006), 71-80.

\bibitem{tsal} C. Tsallis, Possible generalization of Boltzmann-Gibbs statistics,
\emph{J.Stat.Phys}, \textbf{52} (1988), 479-489;
http://tsallis.cat.cbpf.br/biblio.htm.

\bibitem{abr} M. Abramowitz and I.A. Stegun (eds), \emph{Handbook of mathematical
functions with formulas, graphs and mathematical tables}, U. S.
Dept. of Commerce, National Bureau of Standards (1976).

\bibitem{kal74} V.P. Kalashnikov, Interaction of conduction electrons with
an external electromagnetic in the gage-invariant theory of
combined resonance, \emph{Theoretical and Mathematical Physics},
\textbf{18}, (1974), 76-84.

\bibitem{rep81} G. R\"{o}pke, Electrical conductivity of a system of localized
and delocalized electrons, \emph{Theoretical and Mathematical
Physics}, \textbf{46}, (1981), 184-190.

\bibitem{rep91a} V. B. Bobrov, R. Redmer, G. R\"{o}pke, S. A. Triger,
Distribution function and conductivity of a system of charged
particles in linear response theory: Kubo theory and the
nonequilibrium statistical operator method, \emph{Theoretical and
Mathematical Physics}, \textbf{86}, (1991), 207-215.

\bibitem{bobr} V. B. Bobrov, R. Redmer, G. R\"{o}pke, S. A. Triger,
Distribution function and conductivity of a system of charged
particles in linear response theory: the nonequilibrium
statistical operator method and the kinetic equation method,
\emph{Theoretical and Mathematical Physics}, \textbf{86}, (1991),
293-302.

\bibitem{mc1} J.A. McLennan, Nonlinear Effects in Transport
Theory, \emph{Phys. Fluids}, \textbf{4}, (1961), 1319-1324.

\bibitem{mc2} J.A. McLennan, The Formal Statistical Theory of Transport Processes,
\emph{Advan. Chem. Phys.}, \textbf{5}, (1963), 261-317.

\bibitem{ryDr} V.V. Ryazanov, Statistical Theory of Evaporation and Condensation Processes
in Liquid Droplets, \emph{Colloid Journal}, \textbf{68}, (2006),
217-227.

\bibitem{ryAt} V.V. Ryazanov, Distribution of energy of neutrons in the nuclear reactor
in view of finiteness of their lifetime, \emph{Atomnaya energiya
(Atomic energy)}, \textbf{99}, (2005), 348-357.

\bibitem{plas} J-H. Scho\"{e}nfeldt, N. Jimenez, A.R. Plastino, A. Plastino, M.
Casas, Maximum entropy principle and classical evolution equations
with source terms, \emph{Physica A}, \textbf{374}, (2007),
573-584.

\end{thebibliography}
\end{document}